# Improving post-operative discharge destination prediction of geriatric patients with generative data augmentation

Pegah Golchian[1,2], Pauline Maier[3], Thomas Kocar[3], Marvin N. Wright[1,2]

1 Leibniz Institute for Prevention Research and Epidemiology – BIPS, Bremen, Germany

2 Faculty of Mathematics and Computer Science, University of Bremen, Bremen, Germany

3 Institute for Geriatric Research Ulm, Ulm University Medical Center, Ulm, Germany

Correspondence:

Achterstraße 30, 28359 Bremen, Germany

E-Mail: wright@leibniz-bips.de

**Abstract**

*Background*: Data scarcity challenges the development and implementation of innovative healthcare solutions. In geriatrics, fall-related injuries are a major cause of hospitalization, functional decline, and mortality in older adults. Optimizing post-operative discharge planning can mitigate these outcomes, but limited data hinders predictive model development.

*Objectives:* This study aimed to investigate whether generative machine learning based data augmentation can improve predictive performance for post-operative discharge destination in geriatric patients.

*Methods*: We explored generative machine learning approaches to augment data from the SURGE-Ahead project (Supporting SURgery with Geriatric Co-Management and AI), an initiative addressing geriatric perioperative care. Data from the German geriatric trauma register (AltersTraumaZentrum; ATZ) were incorporated using two strategies: (i) combining SURGE-Ahead and ATZ register data with imputation (ComImp) and (ii) generating synthetic data from SURGE-Ahead alone or combined SURGE-Ahead and the ATZ register datasets with Adversarial random forests (ARF). Predictive models, including multinomial logistic regression, random forest, and a prior-fitted transformer (TabPFN), were trained and evaluated using standard performance metrics: accuracy, area under the receiver operating characteristic curve (ROC AUC), Brier score, and the logistic loss.

*Results*: Random forest and TabPFN performed well (accuracy around 0.84 and AUC around 0.94) and were largely unaffected by augmentation. Logistic regression benefited from augmented data, with predictive performance improving from 0.70 to 0.81 for accuracy and 0.85 to 0.92 for AUC.

*Conclusion*: These results highlight generative data augmentation as a viable approach to enhance simpler predictive models in geriatric care and emphasize the importance of method selection when addressing data scarcity in heterogeneous clinical populations.

## 1  Introduction

The development, evaluation, and implementation of innovative healthcare solutions are limited by the availability of high-quality data, a challenge that affects healthcare systems worldwide[1]. The issue is particularly pronounced in the field of geriatrics, where patient heterogeneity, comorbidities, polypharmacy, and variable functional status make it difficult to generate large and high-quality datasets. In high-income Western nations, demographic aging places an increasing strain on healthcare, social and long-term care systems[2]. Efficient, data-driven approaches leveraging machine learning and artificial intelligence (AI) are needed to optimize patient outcomes and resource utilization. Falls and fall-related injuries illustrate the magnitude of these challenges. In 2021, approximately 45.7 million adults aged 65 and older experienced a fall, representing a 182% increase since 1990, with approximately 555,539 fall-related deaths and substantial associated disability-adjusted life years lost[3]. Cohort studies demonstrate that falls are associated with markedly reduced long-term survival: in a five-year European study, older adults with high fall risk and poor physical performance had only 41% survival[4]. One leverage point for intervention is post-operative discharge planning, as timely, clinically appropriate discharge decisions improve patient outcomes and are more cost-effective than delayed decisions[5,6]. For instance, our SURGE-Ahead project (Supporting SURgery with Geriatric Co-Management and AI) uses a dashboard-style interface that integrates a comprehensive geriatric assessment and predictive algorithms to guide surgeons in optimizing discharge decisions, thereby reducing morbidity, functional decline, and dependency in older patients[7,8]. However, initiatives like SURGE-Ahead are constrained by the limited availability of high-quality datasets.

Given the limited sample size in the SURGE-Ahead study, we opted to utilize orthogeriatric register data, the ATZ register (AltersTraumaZentrum; translates to 'geriatric trauma center'), a readily available dataset from a comparable population, to increase the number of potential training samples. However, this introduces two challenges: Firstly, the sample distributions differed due to different inclusion criteria between the SURGE-Ahead study and the ATZ register data (only trauma patients, patients more frail); secondly, the ATZ register data lacked the extensive range of features observed in the SURGE-Ahead study. The two issues can be approached by *Combine datasets based on Imputation (ComImp)*[9], a method that concatenates the datasets by aligning all features and filling the missing features in each dataset with empty columns. The resulting combined dataset thus contains missing values and can be handled with imputation methods, which fill the missing values with point estimates[10–12]. In their publication, the authors also suggested *Principle Component Analysis CompIMP (PCA-ComImp)*, where dimensionality reduction is applied before the datasets are combined. Another approach to dealing with small data is data augmentation, where synthetic data is used to increase sample size[13,14]. Some synthesizing algorithms that work well on tabular data are, for example, *Adversarial random forest (ARF)*[15] and *Synthetic Populations in R (synthpop)*[16], which are generative models based on tree-based machine learning.

In the present work, we demonstrate how novel generative machine learning methods can leverage data from another dataset for augmentation purposes, particularly addressing the issues of data scarcity and disparate source data. We compare the prediction performance of the discharge destination of models trained on the SURGE-Ahead dataset with models trained on data augmented versions. These include combining the dataset from the ATZ register with imputation and generating synthetic datasets based on SURGE-Ahead alone or SURGE-Ahead and ATZ register data combined.

## 2 Methods

### 2.1 SURGE-Ahead and ATZ register data

The SURGE-Ahead study[7], conducted from February 2023 to March 2024 at Ulm University Medical Center, investigated the congruence between standard surgical discharge decisions and geriatric expert recommendations in 178 patients aged 70 and older undergoing surgery. Participants were selected based on an Identification of Seniors At Risk (ISAR) score of 2 or higher and excluded if they had a short anticipated hospital stay, limited communication abilities, or a life expectancy under 3 months. The dataset includes a comprehensive range of variables encompassing demographic data (age, sex), surgical details (hospital department, emergency status, procedure times, OPS-Code[1]), anamnestic information (medication count, care level, living situation, history of falls), and a plethora of functional assessments, including: Montreal Cognitive Assessment 5-min (Wong et al. 2015), Clinical Frailty Scale (CFS), Barthel Index, Charité Mobility Index (CHARMI), American Society of Anaesthesiologists (ASA) score. The primary outcome focused on the agreement between standard care and geriatrician-recommended discharge destinations (back home, acute geriatric care unit, rehabilitation, nursing home), the latter not shown to the treating clinicians. Each patient received at least one geriatric visit within three days before discharge, with recommendations informed by clinical impressions, geriatric assessments, routine care information, and patient preferences. The definitive ground truth was established in a joint consensus conference by both geriatric experts, incorporating 3-month follow-up data and full chart review. Secondary outcomes, evaluated at 3 and 15 months post-discharge, included changes in care level, institutionalization, mobility, activities of daily living, and hospital readmissions. Of the initial 178 participants, 9 did not reach the primary outcome due to dropout or death, and follow-up data was incomplete for some patients at 3 and 15 months. Ultimately, complete data for comparison between standard care and geriatric recommendations was available for 169 patients, revealing a 73% adherence rate and association between non-adherence and poorer clinical outcomes[7,8,17].

To expand our dataset for developing a discharge destination algorithm, we used a retrospective data collection approach. We retrieved patient data from the ATZ register and supplemented it with additional information from the hospital information system (HIS) of the Ulm University Medical Center and the Agaplesion Bethesda Clinic Ulm. Compared to the SURGE-Ahead data, the ATZ register lacked Barthel Index assessments, with the exception of the Barthel Index at the time of discharge. CHARMI assessments were only present in few instances. The CFS and MoCA assessments were also missing, however, correlated assessments within the same geriatric domain were often present to help imputation efforts. In contrast to the SURGE-Ahead data, discharge decisions were made by a geriatric liaison service, not by team consensus with access to 3-month follow-up data (also see Table 1). For an overview comparing the two datasets, we refer to Table 5 in the Appendix.

---

[1] the *OPS-codes* are categorized into 5 categories: '5-55', '5-79', '5-82', '5-83' and 'other'.

*Table 1 Number of samples that fall into the discharge destination classes within the datasets.*

|  | | Discharge destination | | |
| --- | --- | --- | --- | --- |
| Data | acute geriatric care unit | rehabilitation | back home | nursing home |
| SURGE-Ahead | 75 | 9 | 79 | 6 |
| ATZ register | 46 | 6 | 20 | 10 |
| SURGE-Ahead & ATZ register | 121 | 15 | 99 | 16 |

## 2.2 Data augmentation

We compare the predictive performance of models using the SURGE-Ahead dataset with models using augmented versions of it. Data are either augmented by combining them with ATZ register data by imputation or by generating synthetic data based on SURGE-Ahead or on SURGE-Ahead and ATZ register data combined (see Figure 1). We define a dataset $\mathcal{D} = \{\mathbf{x}_i, y_i\}_{i=1}^n$ with $\mathbf{x} \in X \subseteq \mathbb{R}^p$ elements from a feature space and a target $y \in Y \subseteq \mathbb{R}$ with $n \in \mathbb{N}$ observations. In this context, each observation $i$ (row) refers to a patient, where $p$ features (column) were collected, e.g., Barthel Index (describes activities of daily living), and the discharge destination is noted in $y$. In this supervised task, we aim to learn a functional relationship $\hat{f}: X \to Y$, where we want to predict $Y$, which approximates the true relationship. In this paper, we aim to improve the prediction of the discharge destination, if new patients were to come in, by augmenting the dataset used for learning $\hat{f}$, i.e., training a prediction model. In the following, we will first introduce the *Combine datasets based on Imputation (ComImp)*[9] approach and then *Adversarial random forests (ARF)*[15] for synthesizing datasets.

To combine different datasets with imputation according to ComImp, we concatenate the datasets, ensure the features are in the same order, and fill empty columns in the missing features of the resulting combined dataset. This way, we can treat it as one dataset with missing values and handle this with imputation methods. More precisely, let $\mathcal{D}_1, \dots, \mathcal{D}_k$ be the datasets that we want to combine, which can have different dimensions $p_k$ and $n_k$. Some features are shared among datasets, some are missing. Let $\mathcal{F}_{comb}$ be the union of all the features of the different datasets, where each feature appears once. Rearrange the order of the features of each dataset according to $\mathcal{F}_{comb}$, adding previously nonexistent features with empty columns. Stack the $k > 1$ datasets and receive a complete dataset $\mathcal{D}_{comb}$ with $\mathcal{F}_{comb}$ features and $n = n_1 + \cdots + n_k$ rows. The missing values with the resulting combined dataset can now be treated with imputation methods.

For generating synthetic data, we apply adversarial random forests (ARF)[15], which is a tree-based machine learning algorithm based on generative modeling that offers density estimation and data synthesis. With adversarial learning[18], it recursively uses unsupervised random forests[19]—with altering generator and discriminator steps—to learn the data structure of a dataset. In each iteration, a random forest distinguishes between a synthetic dataset and the real dataset (*discriminator step*). The synthetic dataset in the initial step is simply sampling from the marginals of the real dataset (*generator initial step*). In the next generator steps, the generator improves further and samples from the marginals within the leaves of the previous discriminating random forest, which results in a less naïve synthetic dataset. This procedure continues until the accuracy of the random forest (in the discriminating step) is around 0.5, where it cannot distinguish between the datasets anymore, and we then assume that we could generate a dataset very close to the real one. We then use the last random forest that was used in the generating step as the final ARF. In the learning process, ARF learns feature dependencies with the tree structures, meaning we can assume local independence within the leaves, which allows for local density estimation. For that, in each leaf, we apply a univariate density estimation procedure to each feature. To generate a data point with ARF, we then sample a leaf according to leaf weights (weighted by the share of real data points) and then sample a point from the estimated local density.

With ARF we can perform marginal but also conditional density estimation. We estimate a conditional distribution by filtering out the leaves that fulfill the desired conditions and estimating the distribution with those leaves. This makes ARF a suitable tool for algorithms that need flexible and fast sampling strategies. The original publication uses it for generating synthetic data by sampling from the estimated joint distributions. But it also found its application in interpretable machine learning methods such as counterfactual explanations[20] and conditional feature importance[21], where sampling from the conditional distribution was used. The approximation of the conditional distribution can also be used to impute missing values with MissARF[22]. For that, we condition on the non-missing values and sample from the conditional distribution.

## 2.3 Experiment setup

Our aim is to investigate whether we can improve the model prediction of the SURGE-Ahead dataset through data augmentation. To do this, we compare the performance of prediction models trained on SURGE-Ahead with those trained on the augmented versions of the SURGE-Ahead dataset described in Section 2.2. As prediction models we used a multinomial logistic regression, random forest[23], and TabPFN[24]. We use a 5-fold cross-validation with 100 repetitions and evaluate the performance based on accuracy, operating characteristic curve (ROC AUC), Brier score, and logistic loss. For accuracy, we also consider class-wise accuracy. Finally, we compare the global explanation with *permutation feature importance (PFI)*[23,25] –where the contribution of each feature is measured by permuting the values and examining the increase of the prediction error.

For ComImp, we consider two different imputation approaches. The first one is to consider the SURGE-Ahead dataset and the ATZ register as one dataset and apply imputation on the whole dataset (denoted as *SA+ATZ*). The second one is to use the imputation model of SURGE-Ahead for the ATZ register (denoted as $SA \rightarrow ATZ$). As imputation methods, we use *MeanMode Imputation*, which imputes a missing value in a feature with the mean (for numeric data) or the mode (for categorical data) of that feature, and the novel imputation method MissARF[22] that conditions on the non-missing values and samples from an estimated conditional distribution by ARF[15] (described above). Note that with $SA \rightarrow ATZ$ we use the mean or mode of the SURGE-Ahead dataset to impute the ATZ register, whereas with *SA+ATZ* we use the overall mean or mode. In SA $\rightarrow$ ATZ, the features of the ATZ register are reduced to the number of features in the SURGE-Ahead dataset (31 features). For

SA+ATZ, two features regarding some cognitive tests (CIT and MMSE) were added, since they might help the imputation of the dementia values in the ATZ register (33 features).

We apply ARF on the SURGE-Ahead dataset and SURGE-Ahead combined with the ATZ register with the size $n_{synth} \in \{82,338\}$, the same size as the ATZ register and double the size of the SURGE-Ahead dataset.[2] Since the synthetic data has the same data structure as the datasets, they can be stacked into one dataset. The resulting combined dataset is without missing values, where the training data is imputed with MissARF, and for the synthesized data, we choose the option of returning complete data as well. To disentangle sample size effects from information introduced by the ATZ register, we additionally vary the number of synthetic observations over a wider range ($n_{synth} \in [50,500]$) for the logistic regression. An overview of the experiment setup is found in Table 2 and Figure 1.

In general, the imputation model of the training dataset is applied to the test dataset. The test dataset only includes data from SURGE-Ahead. An additional experiment design is that we keep the feature that indicates the source of the data (SA, ATZ) for the imputation model, but leave it out for the prediction model. Furthermore, we keep the outcome variable $y$ in the imputation model to comply with the synthesizing approach, which also includes $y$, and to improve the imputation further.

*Table 2 Description of the augmentation methods used in the experiments.*

| Methods | Description |
|---|---|
| MeanMode | Imputes a missing value in a feature with the mean or the mode of that feature. |
| MissARF | Imputes a value by conditioning on the non-missing values of the observation and sampling from the estimated conditional distribution by ARF. |
| Synth_82 | Creates synthetic data of size 82 with ARF. |
| Synth_338 | Creates synthetic data of size 338 with ARF. |
| SA+ATZ | The SURGE-Ahead dataset and the ATZ register are combined with imputation (MeanMode or MissARF) by concatenating the datasets and considering them as one dataset. The feature size is 33–all the features of SURGE-Ahead and two additional features of the ATZ register. Synth_82 and Synth_338 create synthetic data based on SA+ATZ and return synthetic data without missing values. The synthetic data is then concatenated with the imputed dataset of SA+ATZ with MissARF. |
| SA → ATZ | The SURGE-Ahead dataset and the ATZ register are combined with imputation (MeanMode or MissARF) as follows: First, impute the SURGE-Ahead data and then use that imputation model to also impute the ATZ register. The feature size is 31– only the features of SURGE-Ahead data. |

---

[2] We chose the size of the ATZ register to compare imputation-based augmentation with generative augmentation on the same sample size.

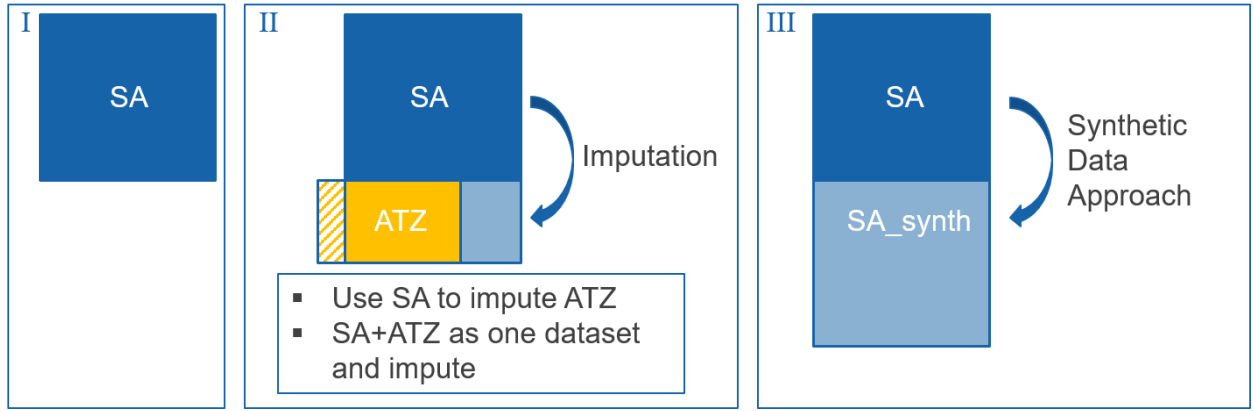

*Figure 1 We compare the predictive performance of the different models using I) the SURGE-Ahead data (SA) as the baseline and II-III) the data augmented versions. These are: II) data augmentation with imputation. Here, we extend the dataset by combining SURGE-Ahead data (SA) and the ATZ register data (ATZ) with imputation (ComImp) with MeanMode imputation or MissARF. We consider two imputation approaches: SA+ATZ, where the dataset is treated as one dataset. Or SA → ATZ, where we apply the learned imputation model of the SURGE-Ahead data (SA) to impute the ATZ register. III) Data augmentation with synthetic data. Here, we learn the data structure of the SURGE-Ahead data (SA) or SA+ATZ and create synthetic data with ARF. We consider sample sizes 82 and 338.*

## 3 Results

### 3.1 Performance comparison

Applying the models on the different datasets, the random forest achieves the best results on average in terms of both accuracy (0.85) and AUC (0.95), followed closely by TabPFN, which performs slightly worse by 0.01 (Figure 2 and Table 6, Table 11 in the Appendix). For both models, the results remain almost identical between the SURGE-Ahead data and the augmented datasets, with only negligible improvements (0.01) observed with MissARF for SA+ATZ. For TabPFN, synthesizing data by 82 extra samples also leads to an improvement; however, SA → ATZ leads to a decrease in performance with MeanMode imputation. However, we observe substantial differences in the logistic regression (Table 3 and Figure 2). Generally, the logistic regression performs rather poorly on just the SURGE-Ahead data with the classical imputation methods, with an accuracy of 0.70 and an AUC of 0.85. Augmenting with the ATZ register improves the performance in both cases, mostly for MeanMode Imputation by up to 0.03. Using synthetic data improves the performance further. With 338 additional synthetic data points in the SURGE-Ahead data, the performance drastically improves to an accuracy of 0.79 and an AUC of 0.90. Combining this with the ATZ register improves this further by 0.02 to an accuracy of 0.81 and an AUC of 0.92, which then comes closer to the performance of the random forest and TabPFN. When we apply a larger range of sample sizes from 50 to 500 for the logistic regression with the Surge-Ahead data and SA+ATZ, the logistic regression improves further with more synthetic data (Figure 5 in the Appendix) and maintains the same trend that adding the ATZ register performs better.

While the overall accuracy of the models is within an acceptable range, the class-wise accuracy (Table 7-10 in the Appendix) shows that the larger classes *back home* and *acute geriatric care unit* achieve high accuracy, while the smaller classes *rehabilitation* and *nursing home* are partially ignored by the models. For class *back home*, the random forest performs best with an accuracy of 0.97 followed by TabPFN with 0.95, with the augmented datasets. As before, data augmentation for the random forest leads to only minor improvements, whereas for TabPFN it is more pronounced for the SURGE-Ahead data by up to 0.03 with synthesizing the data with 338 points. For the logistic regression, the accuracy improves with synthetic data of size 338 from 0.75 (MeanMode) to 0.9 for the SURGE-Ahead data. Applying ComImp generally performs worse here. For *predicting acute geriatric care unit*, the accuracy is generally lower with the random forest at 0.89 and TabPFN at 0.88, and performs best in SA+ATZ. In contrast to before, the logistic regression now performs better when applying ComImp and improves further with additional synthetic data of size 338, i.e., from 0.79 to 0.84 for SA+ATZ and from 0.72 to 0.79 with the SURGE-Ahead data. For *nursing home*, the accuracy of the random forest is very close to zero. TabPFN also has an accuracy $\leq 0.50$ and gets smaller with synthetic data. The logistic regression is $\geq 0.50$ with the SURGE-Ahead data and falls below that with 338 additional synthetic data points. ComImp improves the performance by almost 0.2. It has its best performance at 0.84 in the case of SA $\rightarrow$ ATZ with MissARF. For geriatric *rehabilitation* all models have an accuracy close to zero, with the logistic regression having its highest value at 0.25.

For Brier Score and the logistic loss, we notice a similar pattern as observed for the accuracy and AUC (Figure 4 in the Appendix). The random forest and TabPFN perform well with little effect of data augmentation, and the performance of logistic regression improves with data augmentation, especially with the application of synthetic data.

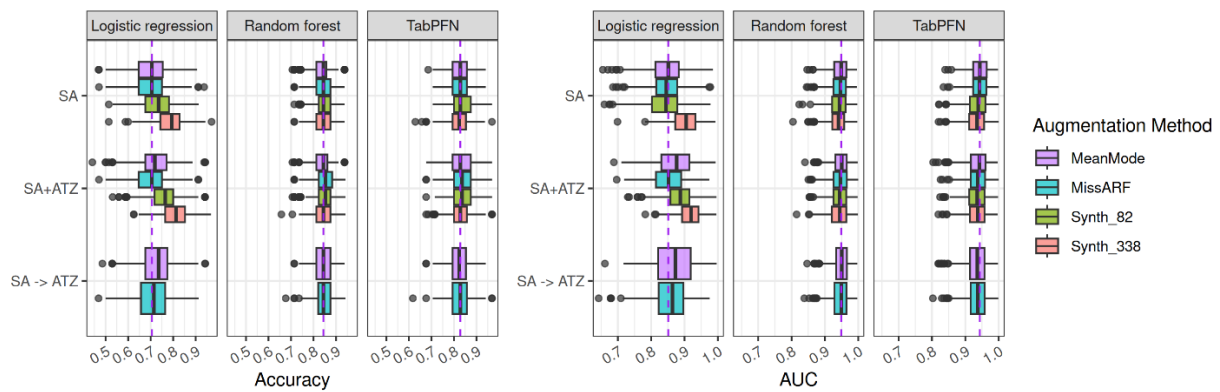

*Figure 2 Accuracy and AUC for logistic regression, random forest, and TabPFN fitted on the SURGE-Ahead data (SA) and its augmented versions of SA. Augmentations include combining SA and the ATZ register with imputation (SA+ATZ, SA $\rightarrow$ ATZ using MeanMode or MissARF) and synthetic data augmentation with 82 (Synth_82) or 338 (Synth_338) additional samples generated with ARF from SA or SA+ATZ. The boxplots are plotted over the cross-validation replicates.*

*Table 3 Accuracy and AUC of a multinomial logistic regression fitted on SURGE-Ahead data (SA) and its augmented versions of SA. Augmentations include combining SA and the ATZ register with imputation (SA+ATZ, SA → ATZ using MeanMode or MissARF) and synthetic data augmentation with 82 (Synth_82) or 338 (Synth_338) additional samples generated with ARF from SA or SA+ATZ. Shown are aggregated results and the standard deviation (SD) over 5 folds and 100 repetitions.*

|  | Accuracy | | | AUC | | |
|---|---|---|---|---|---|---|
| Method | SA [mean (SD)] | SA + ATZ [mean (SD)] | SA → ATZ [mean(SD)] | SA [mean (SD)] | SA + ATZ [mean (SD)] | SA → ATZ [mean(SD)] |
| MeanMode | 0.70 (0.08) | 0.72 (0.08) | 0.73 (0.08) | 0.85 (0.05) | 0.87 (0.06) | 0.87 (0.06) |
| MissARF | 0.70 (0.07) | 0.70 (0.08) | 0.71 (0.08) | 0.85 (0.05) | 0.85 (0.05) | 0.86 (0.05) |
| Synth_82 | 0.73 (0.08) | 0.76 (0.07) | - | 0.84 (0.06) | 0.88 (0.04) | - |
| Synth_338 | **0.79 (0.06)** | **0.81 (0.06)** | - | 0.90 (0.04) | **0.92 (0.04)** | - |

## 3.2 Interpretability

In the following, we interpret the logistic regression model for the best case setting, SA+ATZ with 338 additional synthetic data points (see Table 3). For the logistic regression, *acute geriatric care unit* was used as a reference class. Table 4 shows the odds ratio of the nine most important features according to the PFI values on the test dataset (Figure 3) of the three non-reference discharge categories. The *Barthel index after discharge* is the most important feature. The other variables, which include *CHARMI*, *age* and *previous care situation*, follow with similar importance values.

Higher functional assessment scores in both the *Barthel Index* and the *CHARMI score* were associated with increased odds of discharge *back home* or to *rehabilitation* compared to *acute geriatric care unit* (Table 3). The *Barthel Index at discharge* showed the most stable effect, with each additional point increasing the odds of these destinations by roughly one to three percent, while early postoperative assessments showed the same direction but greater variability. The *previous care situation* was one of the strongest predictors: patients who had lived in a nursing home prior to admission had markedly higher odds of being discharged back to a *nursing home*, whereas those previously living at home had higher odds of returning *back home*. *Age* showed a modest effect, with younger patients more likely to be discharged *back home*. Being treated on a *UCH* or *URO hospital department* was associated with substantially lower odds of discharge to a *nursing home*, although estimates for the smaller classes were less stable. The *OPS codes* did not show consistent effects, but procedures in the *5–83 category*, which relate to spine surgery, tended to be associated with increased odds of discharge to geriatric *rehabilitation*, again with wide variability. In general, *nursing home* and *rehabilitation* are less stable and have a more unreliable interpretation because of their small sample size.

To assess which features the model uses to generalize, we applied PFI on the test data set for each setting. In Figure 6 - 8 in the Appendix, the PFI values for the different models are visualized for the union of the top three features in each setting. In general, the *Barthel Index at discharge* is by far the most important feature and is followed by *Barthel Index post-OP day 3*. The rest of the features are distributed almost equally low. However, for the logistic regression, the ranking of the two features changes with combining the SURGE-Ahead dataset with the ATZ register with imputation with MissARF and MeanMode imputation, where *Barthel Index at discharge* gets less important and is ranked second, closely followed by the *Clinical Fratility Score (CFS)*, which results in increased PFI values. Furthermore, ComImp with MeanMode for TabPFN leads to *previous care situation* becoming more important and closer to the *Barthel Index post-OP day 3*. For logistic regression and TabPFN, we observe that with more synthetic data, previously important features become less important, leading to PFI values becoming more evenly low.

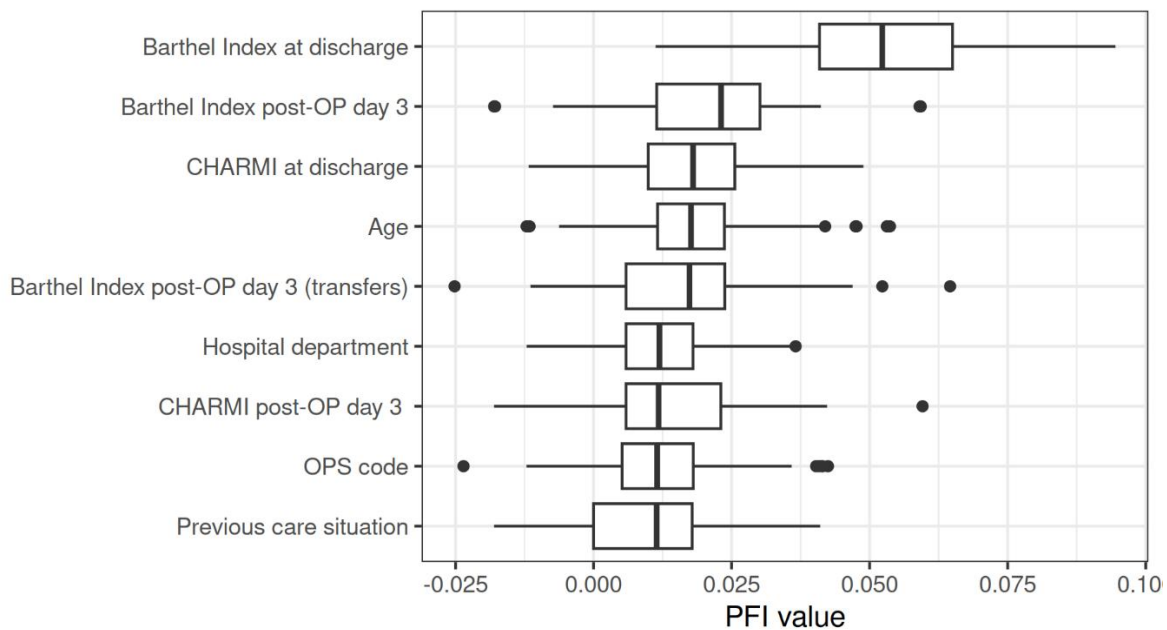

*Figure 3 Top nine features regarding the PFI for the test data of the logistic regression of the case SA+ATZ with 338 additional synthetic data points. The boxplots are plotted over the 100 repetitions (each averaged over 5 folds).*

*Table 4 Odds ratio (OR) from the multinomial logistic regression for the case SA+ATZ with 338 additional synthetic data points, using acute geriatric care unit as the reference class. Shown are the top nine features according to PFI on the test data set. Median values and the 2.5 and 97.5 percent quantiles are calculated across the 100 repetitions (each averaged over 5 folds).*

| Feature | back home | nursing home | rehabilitation |
|---|---|---|---|
| | Median OR (2.5%–97.5% quantile interval) | | |
| Barthel Index at discharge | 1.03 (1.02, 1.03) | 1.01 (1.00, 1.03) | 1.03 (1.02, 1.04) |
| Barthel Index post-OP day 3 | 1.01 (1.01, 1.02) | 1.00 (0.98, 1.02) | 1.01 (1.00, 1.03) |
| CHARMI at discharge | 1.12 (1.06, 1.19) | 1.00 (0.86, 1.16) | 1.10 (1.01, 1.21) |
| Age | 0.95 (0.93, 0.97) | 0.99 (0.95, 1.03) | 1.01 (0.98, 1.04) |
| Barthel Index post-OP day 3 (transfers) | 1.05 (1.01, 1.07) | 1.02 (0.93, 1.09) | 1.08 (1.03, 1.14) |
| Hospital department (reference: AVC) | | | |
|   UCH | 0.66 (0.46, 1.09) | 0.16 (0.07, 0.42) | 2.95 (0.83, 13.81) |
|   URO | 0.74 (0.43, 1.32) | 0.03 (0.00, 0.35) | 0.40 (0.03, 3.25) |
| CHARMI post-OP day 3 | 1.07 (1.01, 1.13) | 1.00 (0.86, 1.16) | 1.10 (0.99, 1.23) |
| OPS code (reference: 5-55) | | | |
|   5-79 | 0.75 (0.38, 1.31) | 1.46 (0.30, 10.92) | 1.82 (0.44, 5.99) |
|   5-82 | 0.86 (0.42, 1.44) | 2.19 (0.41, 14.88) | 2.87 (0.71, 10.96) |
|   5-83 | 0.70 (0.33, 1.39) | 0.25 (0.02, 5.84) | 5.07 (1.12, 17.52) |
|   other | 1.17 (0.62, 1.96) | 2.52 (0.43, 16.84) | 1.64 (0.32, 7.48) |
| Previous care (reference: assisted living) | | | |
|   nursing home | 1.31 (0.70, 2.33) | 16.25 (6.11, 97.47) | 0.56 (0.05, 1.83) |
|   home with help | 1.70 (1.18, 2.41) | 1.80 (0.72, 10.31) | 1.28 (0.76, 2.59) |
|   home without help | 1.70 (0.99, 2.62) | 1.92 (0.24, 10.80) | 0.72 (0.30, 1.77) |

## 4 Discussion

We have shown that the prediction of the post-operative discharge destination of geriatric patients can be improved with data augmentation. The improvement was substantial for the logistic regression, while random forest and TabPFN benefited only slightly or not at all from the augmentation. For the logistic regression, the setting of combining the SURGE-Ahead dataset with the ATZ register and adding 338 synthetic data points with ARF performed best (accuracy 0.81, AUC 0.92). Both accuracy and AUC could be improved by 0.1 compared to using SURGE-Ahead data alone.

For the logistic regression, the synthetic data based on the ATZ register combined with the SURGE-Ahead dataset (SA+ATZ) performs better than the synthetic data using the SURGE-Ahead dataset alone (Figure 5 in the Appendix). This suggests that the logistic regression does not only benefit from a larger sample size, but also leverages information from the ATZ register.

Considering the class-wise accuracy, we noticed that all models have a low accuracy for the two classes *rehabilitation* and *nursing home*, especially random forest and TabPFN. However, these classes have very small sizes in comparison (see Table 1), so that the models optimize overall performance by focusing on the larger classes and thereby tend to ignore the small classes. In future work, this problem of unbalanced data could be addressed[26,27]. Ultimately, miscalibration due to differences in SURGE-Ahead versus ATZ register distributions could be corrected for by calibration methods.

Adding the ATZ register with the imputation methods (ComImp approach), the performance for *back home* decreases for the logistic regression model, while it increases for the other classes. Furthermore, SA → ATZ works better in predicting *back home* than SA+ATZ, whereas for the *acute geriatric care unit* it is the other way around. These effects can be explained as follows: With the ATZ register we gain more information for patients who are sent to *acute geriatric care unit* and we get fewer observations for patients going *back home*. Since SA → ATZ applies the imputation model of the SURGE-Ahead data to impute the ATZ register, it leverages information from the SURGE-Ahead data. Since we are testing only on the dataset retrieved from the SURGE-Ahead data, information that benefits the SURGE-Ahead data for a specific class has an advantage for the prediction. If the ATZ register has information for a specific class, for example, the *acute geriatric care unit*, SA+ATZ has an advantage. Another possible reason could be that the logistic regression performs better with SA → ATZ due to two fewer features (see Section 2.3). In the overall accuracy, the logistic regression tends to perform better in the case of SA → ATZ, and TabPFN and random forest in the case of SA+ATZ. This again can be explained by the logistic regression benefiting from fewer features. However, based on the results, the difference between these two approaches is small. In future work, this effect can be further examined, and the logistic regression could potentially be improved by prior feature selection.

The datasets contain highly correlated features, for example, the assessment of the *Barthel index* at different time points. This affects not only the performance and interpretation of the logistic regression but also the feature importance of PFI. PFI faces extrapolation issues and can create unrealistic observations, which leads to bias. Furthermore, it can decrease the importance by sharing it with the other features[28]. Other feature importance measures could be considered, for example, conditional feature importance like CPI[29] or cARFi[21]. Despite these limitations, the PFI results for the logistic regression model (Figure 3) align well with established predictors of appropriate discharge destinations for older inpatients [30].

Since 5-fold cross-validation was used for model evaluation, model uncertainty is underestimated[31]. In addition, single imputation was used in our experiments, which improves predictive performance but does not consider imputation uncertainty. The datasets exhibit varying patterns of missing data, ranging from completely random to non-random-driven by factors such as patient fitness, cognitive impairment, and differences in study inclusion criteria. Furthermore, the ComImp method itself suggests a missing at random pattern, due to missingess in features dependent on the data source. Applying single imputation can therefore lead to biased estimates[11]. The results of interpretable machine learning methods can change when learner[32] and imputation uncertainty[33–35] are not considered. Molnar et al.[32] showed approximately valid confidence intervals for the learner uncertainty by refitting the model on 15 bootstrapped or subsampled datasets with adjusted variance[31]. Golchian and Wright[35] suggest including the imputation uncertainty on top of the learner uncertainty in the explanations by applying multiple imputation methods. In future work, when applying interpretable machine learning methods, learner and imputation uncertainty could be included with bootstrap and the application of multiple imputation methods such as MICE[36] or MissARF[22].

For healthcare providers seeking to understand a machine learning model, explaining individual predictions (local explanations) is often more valuable than explaining the model's overall behavior. A common approach is providing Shapley values[37], which could be enhanced in the future to incorporate imputation uncertainty. Alternatively, intrinsically explainable models, such as logistic regression, are frequently favored due to their simplicity and robustness. Data augmentation techniques designed specifically for these models may not only improve performance but also provide intelligent imputation solutions for missing data. While the final output adheres to the decision function of the intrinsically explainable model, sophisticated imputation and augmentation methods in the background can extract the maximum information from the available dataset.

Finally, we emphasize the proof-of-concept nature of this study. A significant challenge in implementing large-scale machine learning solutions in healthcare lies in data heterogeneity, stemming from variations in patient populations[1] and technical factors[38]. Obtaining both marginal and conditional density estimation using ARF is a valuable asset in addressing this issue. In geriatrics specifically, numerous assessments target the same clinical domain but are often difficult to translate directly[39]. Future work could extend the methods presented in this proof-of-concept study to large-scale, transnational datasets, enabling on-demand imputation with uncertainty through conditional density estimation. Ultimately, a machine learning model trained on one set of features could potentially be applied to a different set of features, fostering international collaboration and facilitating the implementation of AI solutions in healthcare.

## 5  Conclusion

In this study, we applied novel generative machine learning models to augment the data of the SURGE-Ahead dataset to improve the prediction performance of the post-operative discharge destinations. We combined the dataset with the ATZ register with imputation (ComImp), i.e., MeanMode and MissARF, and compared it with synthesizing data with ARF. Our results show that for the machine learning models, the random forest and TabPFN, augmenting the data only led to minor improvements, while for the logistic regression, the performance improved more with synthesizing data. This allows the application of an intrinsically explainable model, which performance can be further improved. In future work, unbalanced data and highly correlated features have to be addressed. For the latter, we recommend feature selection based on the new results. When applying interpretable machine learning methods, we recommend the application of bootstrap and multiple imputation methods to consider learner and imputation uncertainty. However, the effect of data synthesis on the interpretability has to be further examined. In future work, the local explanations and dealing with new patients can be studied. Finally, future work could extend the methods presented in this proof-of-concept study to large-scale, transnational datasets.

### Acknowledgments

PG and MNW were supported by the German Research Foundation (DFG), Grant Number 437611051. TDK was supported as part of the SURGE-Ahead project, for which funding was granted by the German Federal Ministry of Education and Research, Grant numbers 01GY2101 and 01GY2404. We thank Prof. Dr. Hans Kestler and Prof. Dr. Michael Denkinger for the valuable discussions.

**Ethics and Consent Statement**

The observational and AI development study of the SURGE-Ahead project was conducted following ethical guidelines set by the Declaration of Helsinki. It was approved by the University of Ulm's Ethical Committee with reference number # 310/22-Sta. All participants provided their written informed consent.

**Conflict of Interest**

None declared.

**Data and Code Availability**

The code can be found in https://github.com/PegahGolchian/SurgeAHEAD.git and structurally similar open-source dataset in https://github.com/IfGF-UUlm/SA_COC/blob/main/coc_data.csv.

**Abbreviations**

| Abbreviation | Description |
| --- | --- |
| AI | Artificial intelligence |
| ARF | Adversarial random forest |
| ATZ register | AltersTraumaZentrum; translates to 'geriatric trauma center' |
| AUC | area under the receiver operating characteristic curve |
| ComImp | Combine datasets based on Imputation |
| MeanMode | Imputes a missing value in a feature with the mean or the mode of that feature. |
| MissARF | Imputes a value by conditioning on the non-missing values of the observation and sampling from the estimated conditional distribution by ARF. |
| PFI | Permutation feature importance |
| SA+ATZ | The SURGE-Ahead dataset and the ATZ register are combined with imputation (MeanMode or MissARF) by concatenating the datasets and considering them as one dataset. The feature size is 33–all the features of SURGE-Ahead and two additional features of the ATZ register. Synth_82 and Synth_338 create synthetic data based on SA+ATZ and return synthetic data without missing values. The synthetic data is then concatenated with the imputed dataset of SA+ATZ with MissARF. |
| SA → ATZ | The SURGE-Ahead dataset and the ATZ register are combined with imputation (MeanMode or MissARF) as follows: First, impute the SURGE-Ahead data and then use that imputation model to also impute the ATZ register. The feature size is 31–only the features of SURGE-Ahead data. |
| SURGE-Ahead | Supporting SURgery with Geriatric Co-Management and AI |
| Synth_82 | Creates synthetic data of size 82 with ARF. |
| Synth_338 | Creates synthetic data of size 338 with ARF. |

## References

1. Ibrahim H, Liu X, Zariffa N, Morris AD, Denniston AK. Health data poverty: an assailable barrier to equitable digital health care. Lancet Digit Health 2021 Apr;3(4):e260–e265. doi: 10.1016/S2589-7500(20)30317-4

2. World Health Organization. World Report on Ageing and Health. World Health Organization; 2015.

3. Chen Y, Dai F, Huang S, Qi D, Peng C, Zhang A, Wang Y, Gu Y, Guo J. Global, regional, and national burden of falls among older adults: findings from the Global Burden of Disease Study 2021 and Projections to 2040. Npj Aging 2025 Oct 9;11(1):85. doi: 10.1038/s41514-025-00275-4

4. Salis F, Mandas A. Physical Performance and Falling Risk Are Associated with Five-Year Mortality in Older Adults: An Observational Cohort Study. Med Kaunas Lith 2023 May 17;59(5):964. PMID:37241196

5. Gonçalves-Bradley DC, Lannin NA, Clemson L, Cameron ID, Shepperd S. Discharge planning from hospital. Cochrane Database Syst Rev 2022 Feb 24;2(2):CD000313. PMID:35199849

6. Nicolet A, Al-Gobari M, Perraudin C, Wagner J, Peytremann-Bridevaux I, Marti J. Association between continuity of care (COC), healthcare use and costs: what can we learn from claims data? A rapid review. BMC Health Serv Res 2022 May 16;22(1):658. PMID:35578226

7. Leinert C, Fotteler M, Kocar TD, Dallmeier D, Kestler HA, Wolf D, Gebhard F, Uihlein A, Steger F, Kilian R, Mueller-Stierlin AS, Michalski CW, Mihaljevic A, Bolenz C, Zengerling F, Leinert E, Schütze S, Hoffmann TK, Onder G, Andersen-Ranberg K, O'Neill D, Wehling M, Schobel J, Swoboda W, Denkinger M, SURGE-Ahead Study Group. Supporting SURgery with GEriatric Co-Management and AI (SURGE-Ahead): A study protocol for the development of a digital geriatrician. PLoS One 2023;18(6):e0287230. PMID:37327245

8. Kocar TD, Wolf P, Leinert C, Brefka S, Fotteler ML, Uihlein A, Wezel F, Wehling M, Rahbari N, Kestler H, Gebhard F, Dallmeier D, Denkinger M. SURGE-ahead postoperative delirium prediction: external validation and open-source library. Eur Geriatr Med 2025 Mar 10; doi: 10.1007/s41999-025-01180-5

9. Nguyen T, Khadka R, Phan N, Yazidi A, Halvorsen P, Riegler MA. Combining datasets to improve model fitting. 2023 Int Jt Conf Neural Netw IJCNN IEEE; 2023. p. 1–9.

10. Molenberghs G, Fitzmaurice G, Kenward MG, Tsiatis A, Verbeke G. Handbook of Missing Data Methodology. CRC Press; 2015.

11. Van Buuren S. Flexible Imputation of Missing Data. CRC Press; 2018.

12. Stekhoven DJ, Bühlmann P. MissForest—non-parametric missing value imputation for mixed-type data. Bioinformatics 2012;28(1):112–118.

13. Moreno-Barea FJ, Jerez JM, Franco L. Improving classification accuracy using data augmentation on small data sets. Expert Syst Appl 2020;161:113696.

14. Pezoulas VC, Zaridis DI, Mylona E, Androutsos C, Apostolidis K, Tachos NS, Fotiadis DI. Synthetic data generation methods in healthcare: A review on open-source tools and methods. Comput Struct Biotechnol J 2024;23:2892–2910.

15. Watson DS, Blesch K, Kapar J, Wright MN. Adversarial random forests for density estimation and generative modeling. Proc 26th Int Conf Artif Intell Stat PMLR; 2023. p. 5357–5375.

16. Nowok B, Raab GM, Dibben C. synthpop: Bespoke creation of synthetic data in R. J Stat Softw 2016;74:1–26.

17. Leinert C, Brefka S, Fotteler ML, Müller-Stierlin AS, Gebhard F, Rahbari N, Bolenz C, Kestler H, Dallmeier D, Denkinger M, Kocar TD. Standard-of-Care vs. Expert-Recommended Discharge Destinations for Geriatric Surgical Inpatients: A Prospective Observational Cohort Study. Eur Geriatr Med 2025;in press.

18. Goodfellow I, Pouget-Abadie J, Mirza M, Xu B, Warde-Farley D, Ozair S, Courville A, Bengio Y. Generative adversarial nets. Adv Neural Inf Process Syst 2014.

19. Shi T, Horvath S. Unsupervised learning with random forest predictors. J Comput Graph Stat 2006;15(1):118–138.

20. Dandl S, Blesch K, Freiesleben T, König G, Kapar J, Bischl B, Wright MN. CountARFactuals–generating plausible model-agnostic counterfactual explanations with adversarial random forests. World Conf Explain Artif Intell Springer; 2024. p. 85–107.

21. Blesch K, Koenen N, Kapar J, Golchian P, Burk L, Loecher M, Wright MN. Conditional feature importance with generative modeling using adversarial random forests. Proc AAAI Conf Artif Intell 2025. p. 15596–15604.

22. Golchian P, Kapar J, Watson DS, Wright MN. Missing Value Imputation With Adversarial Random Forests—MissARF. Stat Med 2026;45(3–5):e70379. doi: 10.1002/sim.70379

23. Breiman L. Random forests. Mach Learn 2001;45(1):5–32.

24. Hollmann N, Müller S, Eggensperger K, Hutter F. TabPFN: A Transformer That Solves Small Tabular Classification Problems in a Second. NeurIPS 2022 First Table Represent Workshop 2022.

25. Fisher A, Rudin C, Dominici F. All Models are Wrong, but Many are Useful: Learning a Variable's Importance by Studying an Entire Class of Prediction Models Simultaneously. J Mach Learn Res 2019;20(177):1–81.

26. Salmi M, Atif D, Oliva D, Abraham A, Ventura S. Handling imbalanced medical datasets: review of a decade of research. Artif Intell Rev 2024;57(10):273.

27. Altalhan M, Algarni A, Alouane MT-H. Imbalanced data problem in machine learning: A review. IEEE Access 2025;

28. Molnar C. Interpretable Machine Learning: A Guide for Making Black Box Models Explainable. 3rd ed. 2025. Available from: https://christophm.github.io/interpretable-ml-book

29. Watson DS, Wright MN. Testing conditional independence in supervised learning algorithms. Mach Learn 2021;110(8):2107–2129.

30. Leinert C, Fotteler ML, Kocar TD, Wolf J, Beissel L, Grummich K, Dallmeier D, Denkinger M. Identifying key predictors of appropriate discharge destinations for older inpatients in acute care: A scoping review. Interact J Med Res 2025;14(e76582). doi: 10.2196/76582

31. Nadeau C, Bengio Y. Inference for the Generalization Error. Mach Learn 2003;52:239–281.

32. Molnar C, Freiesleben T, König G, Herbinger J, Reisinger T, Casalicchio G, Wright MN, Bischl B. Relating the partial dependence plot and permutation feature importance to the data generating process. World Conf Explain Artif Intell Springer; 2023. p. 456–479.

33. Vo TL, Nguyen T, Hammer HL, Riegler MA, Halvorsen P. Explainability of Machine Learning Models under Missing Data. 2024. Available from: arxiv.org/abs/2407.00411v2

34. Erez IB, Flokstra J, Poel M, van Keulen M. The Impact of Missing Data Imputation on Model Performance and Explainability. BNAICBeNeLearn 2024 Jt Int Sci Conf AI Mach Learn 2024.

35. Golchian P, Wright MN. Imputation Uncertainty in Interpretable Machine Learning Methods. 2025. Available from: arxiv.org/abs/2512.17689v1

36. Buuren S van, Groothuis-Oudshoorn K. mice: Multivariate imputation by chained equations in R. J Stat Softw 2011;45(3):1–67.

37. Shapley LS. A Value for n-Person Games. In: Kuhn HW, Tucker AW, editors. Contrib Theory Games Princeton: Princeton University Press; 1953. p. 307–317.

38. Rosenau L, Behrend P, Wiedekopf J, Gruendner J, Ingenerf J. Uncovering Harmonization Potential in Health Care Data Through Iterative Refinement of Fast Healthcare Interoperability Resources Profiles Based on Retrospective Discrepancy Analysis: Case Study. JMIR Med Inform 2024 July 23;12:e57005. PMID:39042420

39. Brefka S, Dallmeier D, Mühlbauer V, von Arnim CAF, Bollig C, Onder G, Petrovic M, Schönfeldt-Lecuona C, Seibert M, Torbahn G, Voigt-Radloff S, Haefeli WE, Bauer JM,

Denkinger MD, Medication and Quality of Life Research Group. A Proposal for the Retrospective Identification and Categorization of Older People With Functional Impairments in Scientific Studies-Recommendations of the Medication and Quality of Life in Frail Older Persons (MedQoL) Research Group. J Am Med Dir Assoc 2019 Feb;20(2):138–146. PMID:30638832

# Appendix

*Table 5 Dataset characteristics. Mean and standard deviation (SD) are given for continuous and discrete features, frequency and percent for categorical features. Unknown shows the number of missing values.*

| Features | SURGE-Ahead (SA) (N = 169) | | ATZ-register (N = 82) | |
|---|---|---|---|---|
| | mean (SD) or n (p%) | Missings | mean (SD) or n (p%) | Missings |
| **Hospital department** | | | | |
| AVC | 22 (13%) | | - | |
| UCH | 127 (75%) | | 82 (100%) | |
| URO | 20 (12%) | | - | |
| **Sex** | | | | |
| male | 70 (41%) | | 24 (29%) | |
| female | 99 (59%) | | 58 (71%) | |
| **Age (years)** | 80 (6) | | 85 (6) | |
| **Time to OP (hours)** | 3,753 (5,259) | 1 | 2,058 (3,393) | 2 |
| **Cut-to-Suture Time (minutes)** | 90 (67) | 1 | 68 (40) | 4 |
| **Length of stay in intensive care unit (ICU) (minutes)** | 409 (977) | 18 | 100 (56) | 29 |
| **Blood transfusion** | | | | 51 |
| yes | 54 (32%) | | 10 (32%) | |
| no | 115 (68%) | | 21 (68%) | |
| **Social grade level** | | | | 18 |
| 0 | 97 (57%) | | 26 (41%) | |
| 1 | 13 (7.7%) | | 4 (6.3%) | |
| 2 | 32 (19%) | | 15 (23%) | |
| 3 | 22 (13%) | | 15 (23%) | |
| 4 | 4 (2.4%) | | 4 (6.3%) | |
| 5 | 1 (0.6%) | | 0 (0%) | |
| **Previous care situation** | | | | 4 |
| assisted living | 5 (3.0%) | | 11 (14%) | |
| nursing home | 17 (10%) | | 12 (15%) | |
| home with help | 123 (73%) | | 41 (53%) | |
| home without help | 24 (14%) | | 14 (18%) | |
| **Needs help in Activities of Daily Living (ADL)** | | | | 3 |
| yes | 145 (86%) | | 65 (82%) | |
| no | 24 (14%) | | 14 (18%) | |
| **Number of social contacts** | 10 (9) | 3 | - | 82 |
| **Barthel Index pre-illness** | 86 (19) | 1 | - | 82 |
| **Barthel Index pre-OP** | 58 (31) | | - | 82 |
| **Barthel Index post-OP day 1** | 43 (22) | 6 | - | 82 |
| **Barthel Index post-OP day 3** | 52 (24) | 14 | - | 82 |
| **Barthel Index post-OP day 3 subscore ("transfers")** | | 14 | - | 82 |
| 0 | 21 (14%) | | | |
| 5 | 47 (30%) | | - | |
| 10 | 23 (15%) | | - | |
| 15 | 64 (41%) | | - | |
| **Barthel Index at discharge** | 62 (24) | 1 | 40 (18) | 31 |
| **CHARMI pre-illness** | 7.95 (2.66) | 5 | 9.57 (1.24) | 59 |
| **CHARMI post-OP day 1** | 2.59 (2.61) | 5 | - | 82 |
| **CHARMI post-OP day 3** | 4.27 (2.72) | 15 | - | 82 |
| **CHARMI at discharge** | 5.45 (2.77) | 1 | 6.78 (1.59) | 64 |
| **ISAR (score)** | | | | 41 |
| 1 | - | | 3 (7.3%) | |
| 2 | 57 (34%) | | 4 (9.8%) | |
| 3 | 64 (38%) | | 16 (39%) | |
| 4 | 36 (21%) | | 7 (17%) | |
| 5 | 9 (5.3%) | | 10 (24%) | |
| 6 | 3 (1.8%) | | 1 (2.4%) | |
| **Frailty impression** | | | | 16 |
| yes | 81 (48%) | | 43 (65%) | |
| no | 88 (52%) | | 23 (35%) | |
| **Clinical Frailty Scale (score)** | | | | 82 |
| 1 | 4 (2.4%) | | - | |
| 2 | 36 (21%) | | - | |
| 3 | 32 (19%) | | - | |
| 4 | 16 (9.5%) | | - | |
| 5 | 32 (19%) | | - | |
| 6 | 33 (20%) | | - | |
| 7 | 15 (8.9%) | | - | |
| 8 | 1 (0.6%) | | - | |
| **ASA (score)** | | 1 | | 2 |
| 0 | 2 (1.2%) | | 0 (0%) | |
| 1 | 23 (14%) | | 0 (0%) | |
| 2 | 128 (76%) | | 7 (8.8%) | |
| 3 | 15 (8.9%) | | 62 (78%) | |
| 4 | 0 (0%) | | 11 (14%) | |
| **Modified Charlson Comorbidity Index (mCCI) (score)** | | 0 | | 5 |
| 0 | 46 (27%) |  | 8 (10%) |  |
| 1 | 58 (34%) |  | 17 (22%) |  |
| 2 | 49 (29%) |  | 18 (23%) |  |
| 3 | 10 (5.9%) |  | 13 (17%) |  |
| 4 | 6 (3.6%) |  | 12 (16%) |  |
| 5 | 0 (0%) |  | 3 (3.9%) |  |
| 6 | 0 (0%) |  | 5 (6.5%) |  |
| 7 | 0 (0%) |  | 1 (1.3%) |  |
| **Number of medications** | 8.9 (3.9) |  | 6.9 (3.5) | 5 |
| **Montreal Cognitive Assessment (MoCA) 5min (score)** | 21.7 (5.9) | 18 | - | 82 |
| **Dementia** |  |  |  | 31 |
| yes | 19 (11%) |  | 27 (53%) |  |
| no | 150 (89%) |  | 24 (47%) |  |
| **Cognitive Impairment Test (CIT) (score)** | - | 169 | 23.9 (5.9) | 44 |
| **Mini-Mental State Examination (MMSE) (score)** | - | 169 | 21.6 (6.3) | 43 |
| **OPS Code** |  |  |  |  |
| 5-55 | 12 (7.1%) |  | 0 (0%) |  |
| 5-79 | 64 (38%) |  | 47 (57%) |  |
| 5-82 | 28 (17%) |  | 25 (30%) |  |
| 5-83 | 8 (4.7%) |  | 3 (3.7%) |  |
| other | 57 (34%) |  | 7 (8.5%) |  |

*Table 6 Accuracy of the different models fitted on SURGE-Ahead data (SA) and its augmented versions of SA. Augmentations include combining SA and the ATZ register with imputation (SA+ATZ, SA → ATZ using MeanMode or MissARF) and synthetic data augmentation with 82 (Synth_82) or 338 (Synth_338) additional samples generated with ARF from SA or SA+ATZ. Shown are aggregated results and the standard deviation (SD) over 5 folds and 100 repetitions.*

|  | Logistic regression | | | Random forest | | | TabPFN | | |
|---|---|---|---|---|---|---|---|---|---|
| Method | SA | SA + ATZ | SA → ATZ | SA | SA + ATZ | SA → ATZ | SA | SA + ATZ | SA → ATZ |
|  | [mean (SD)] | [mean (SD)] | [mean (SD)] | [mean (SD)] | [mean (SD)] | [mean (SD)] | [mean (SD)] | [mean (SD)] | [mean (SD)] |
| MeanMode | 0.70 (0.08) | 0.72 (0.08) | 0.73 (0.08) | 0.84 (0.04) | 0.84 (0.04) | 0.84 (0.04) | 0.83 (0.05) | 0.83 (0.05) | 0.82 (0.05) |
| MissARF | 0.70 (0.07) | 0.70 (0.08) | 0.71 (0.08) | 0.84 (0.04) | **0.85 (0.04)** | 0.84 (0.04) | 0.83 (0.05) | 0.84 (0.05) | 0.83 (0.05) |
| Synth_82 | 0.73 (0.08) | 0.76 (0.07) | - | 0.84 (0.04) | 0.84 (0.04) | - | 0.84 (0.05) | 0.84 (0.04) | - |
| Synth_338 | **0.79 (0.06)** | **0.81 (0.06)** | - | 0.84 (0.04) | 0.84 (0.04) | - | 0.83 (0.05) | 0.83 (0.05) | - |

Table 7 Class-wise accuracy of class back home of the different models fitted on SURGE-Ahead data (SA) and its augmented versions of SA. Augmentations include combining SA and the ATZ register with imputation (SA+ATZ, SA → ATZ using MeanMode or MissARF) and synthetic data augmentation with 82 (Synth_82) or 338 (Synth_338) additional samples generated with ARF from SA or SA+ATZ. Shown are aggregated results and the standard deviation (SD) over 5 folds and 100 repetitions.

| | Logistic regression | | | Random forest | | | TabPFN | | |
|---|---|---|---|---|---|---|---|---|---|
| Method | SA | SA + ATZ | SA → ATZ | SA | SA + ATZ | SA → ATZ | SA | SA + ATZ | SA → ATZ |
| | [mean (SD)] | [mean (SD)] | [mean(SD)] | [mean (SD)] | [mean (SD)] | [mean(SD)] | [mean (SD)] | [mean (SD)] | [mean(SD)] |
| MeanMode | 0.75 (0.11) | 0.71 (0.12) | 0.73 (0.12) | 0.96 (0.06) | 0.96 (0.05) | 0.95 (0.06) | 0.93 (0.07) | 0.93 (0.07) | 0.92 (0.07) |
| MissARF | 0.75 (0.11) | 0.70 (0.13) | 0.72 (0.12) | 0.96 (0.05) | 0.97 (0.05) | 0.96 (0.05) | 0.92 (0.07) | 0.93 (0.06) | 0.93 (0.07) |
| Synth_82 | 0.80 (0.11) | 0.81 (0.11) | - | 0.96 (0.05) | 0.96 (0.05) | - | 0.94 (0.06) | 0.94 (0.06) | - |
| Synth_338 | 0.90 (0.08) | 0.89 (0.09) | - | 0.97 (0.05) | 0.95 (0.06) | - | 0.95 (0.06) | 0.94 (0.06) | - |

Table 8 Class-wise accuracy of class acute geriatric care unit of the different models fitted on SURGE-Ahead data (SA) and its augmented versions of SA. Augmentations include combining SA and the ATZ register with imputation (SA+ATZ, SA → ATZ using MeanMode or MissARF) and synthetic data augmentation with 82 (Synth_82) or 338 (Synth_338) additional samples generated with ARF from SA or SA+ATZ. Shown are aggregated results and the standard deviation (SD) over 5 folds and 100 repetitions.

| | Logistic regression | | | Random forest | | | TabPFN | | |
|---|---|---|---|---|---|---|---|---|---|
| Method | SA | SA + ATZ | SA → ATZ | SA | SA + ATZ | SA → ATZ | SA | SA + ATZ | SA → ATZ |
| | [mean (SD)] | [mean (SD)] | [mean(SD)] | [mean (SD)] | [mean (SD)] | [mean(SD)] | [mean (SD)] | [mean (SD)] | [mean(SD)] |
| MeanMode | 0.72 (0.11) | 0.79 (0.10) | 0.78 (0.10) | 0.88 (0.08) | 0.88 (0.07) | 0.88 (0.08) | 0.87 (0.08) | 0.87 (0.08) | 0.86 (0.08) |
| MissARF | 0.73 (0.11) | 0.76 (0.11) | 0.75 (0.10) | 0.87 (0.08) | 0.88 (0.08) | 0.88 (0.08) | 0.87 (0.08) | 0.88 (0.08) | 0.85 (0.08) |
| Synth_82 | 0.73 (0.12) | 0.79 (0.10) | - | 0.88 (0.08) | 0.88 (0.08) | - | 0.87 (0.09) | 0.88 (0.08) | - |
| Synth_338 | 0.79 (0.11) | 0.84 (0.09) | - | 0.87 (0.08) | 0.89 (0.07) | - | 0.86 (0.09) | 0.88 (0.08) | - |

Table 9 Class-wise accuracy of class nursing home of the different models fitted on SURGE-Ahead data (SA) and its augmented versions of SA. Augmentations include combining SA and the ATZ register with imputation (SA+ATZ, SA → ATZ using MeanMode or MissARF) and synthetic data augmentation with 82 (Synth_82) or 338 (Synth_338) additional samples generated with ARF from SA or SA+ATZ. Shown are aggregated results and the standard deviation (SD) over 5 folds and 100 repetitions.

| | Logistic regression | | | Random forest | | | TabPFN | | |
|---|---|---|---|---|---|---|---|---|---|
| Method | SA | SA + ATZ | SA → ATZ | SA | SA + ATZ | SA → ATZ | SA | SA + ATZ | SA → ATZ |
| | [mean (SD)] | [mean (SD)] | [mean(SD)] | [mean (SD)] | [mean (SD)] | [mean(SD)] | [mean (SD)] | [mean (SD)] | [mean(SD)] |
| MeanMode | 0.58 (0.47) | 0.76 (0.41) | 0.69 (0.44) | 0.00 (0.04) | 0.00 (0.00) | 0.00 (0.00) | 0.36 (0.46) | 0.46 (0.47) | 0.32 (0.45) |
| MissARF | 0.54 (0.48) | 0.71 (0.43) | 0.84 (0.35) | 0.00 (0.02) | 0.06 (0.24) | 0.07 (0.25) | 0.37 (0.46) | 0.31 (0.44) | 0.50 (0.47) |
| Synth_82 | 0.54 (0.48) | 0.62 (0.46) | - | 0.01 (0.09) | 0.05 (0.21) | - | 0.26 (0.42) | 0.25 (0.41) | - |
| Synth_338 | 0.37 (0.46) | 0.45 (0.47) | - | 0.02 (0.13) | 0.05 (0.20) | - | 0.11 (0.30) | 0.15 (0.34) | - |

Table 10 Class-wise accuracy of class geriatric rehabilitation of the different models fitted on SURGE-Ahead data (SA) and its augmented versions of SA. Augmentations include combining SA and the ATZ register with imputation (SA+ATZ, SA → ATZ using MeanMode or MissARF) and synthetic data augmentation with 82 (Synth_82) or 338 (Synth_338) additional samples generated with ARF from SA or SA+ATZ. Shown are aggregated results and the standard deviation (SD) over 5 folds and 100 repetitions.

| | Logistic regression | | | Random forest | | | TabPFN | | |
|---|---|---|---|---|---|---|---|---|---|
| Method | SA | SA + ATZ | SA → ATZ | SA | SA + ATZ | SA → ATZ | SA | SA + ATZ | SA → ATZ |
| | [mean (SD)] | [mean (SD)] | [mean(SD)] | [mean (SD)] | [mean (SD)] | [mean(SD)] | [mean (SD)] | [mean (SD)] | [mean(SD)] |
| MeanMode | 0.14 (0.47) | 0.23 (0.41) | 0.25 (0.44) | 0.00 (0.04) | 0.00 (0.00) | 0.00 (0.00) | 0.01 (0.46) | 0.00 (0.47) | 0.00 (0.45) |
| MissARF | 0.14 (0.48) | 0.19 (0.43) | 0.24 (0.35) | 0.00 (0.02) | 0.00 (0.24) | 0.00 (0.25) | 0.01 (0.46) | 0.00 (0.44) | 0.00 (0.47) |
| Synth_82 | 0.20 (0.48) | 0.17 (0.46) | - | 0.00 (0.09) | 0.00 (0.21) | - | 0.00 (0.42) | 0.00 (0.41) | - |
| Synth_338 | 0.13 (0.46) | 0.07 (0.47) | - | 0.00 (0.13) | 0.00 (0.20) | - | 0.00 (0.30) | 0.00 (0.34) | - |

*Table 11 AUC of the different models fitted on SURGE-Ahead data (SA) and its augmented versions of SA. Augmentations include combining SA and the ATZ register with imputation (SA+ATZ, SA → ATZ using MeanMode or MissARF) and synthetic data augmentation with 82 (Synth_82) or 338 (Synth_338) additional samples generated with ARF from SA or SA+ATZ. Shown are aggregated results and the standard deviation (SD) over 5 folds and 100 repetitions.*

|  | Logistic regression | | | Random forest | | | TabPFN | | |
|---|---|---|---|---|---|---|---|---|---|
| Method | SA | SA + ATZ | SA → ATZ | SA | SA + ATZ | SA → ATZ | SA | SA + ATZ | SA → ATZ |
|  | [mean (SD)] | [mean (SD)] | [mean(SD)] | [mean (SD)] | [mean (SD)] | [mean(SD)] | [mean (SD)] | [mean (SD)] | [mean(SD)] |
| MeanMode | 0.85 (0.05) | 0.87 (0.06) | 0.87 (0.06) | 0.94 (0.03) | 0.95 (0.03) | 0.95 (0.03) | 0.94 (0.03) | 0.94 (0.03) | 0.93 (0.03) |
| MissARF | 0.85 (0.05) | 0.85 (0.05) | 0.86 (0.05) | 0.94 (0.03) | 0.94 (0.03) | 0.95 (0.03) | 0.94 (0.03) | 0.94 (0.03) | 0.93 (0.03) |
| Synth_82 | 0.84 (0.06) | 0.88 (0.04) | - | 0.94 (0.03) | 0.94 (0.03) | - | 0.93 (0.03) | 0.93 (0.03) | - |
| Synth_338 | 0.90 (0.04) | **0.92 (0.04)** | - | 0.94 (0.03) | 0.94 (0.03) | - | 0.93 (0.03) | 0.93 (0.03) | - |

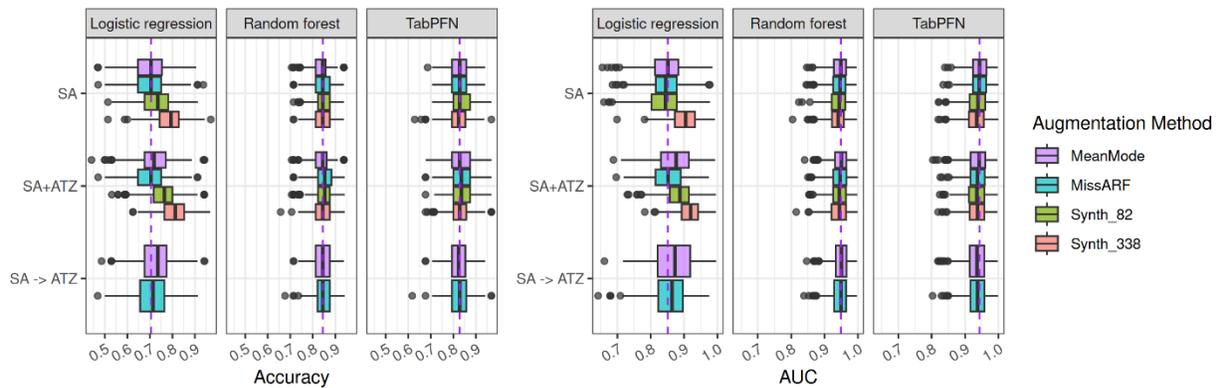

*Figure 4 Brier Score and logistic loss for logistic regression, random forest, and TabPFN fitted on the SURGE-Ahead data (SA) and its augmented versions of SA. Augmentations include combining SA and the ATZ register with imputation (SA+ATZ, SA → ATZ using MeanMode or MissARF) and synthetic data augmentation with 82 (Synth_82) or 338 (Synth_338) additional samples generated with ARF from SA or SA+ATZ. The boxplots are plotted over the cross-validation replicates.*

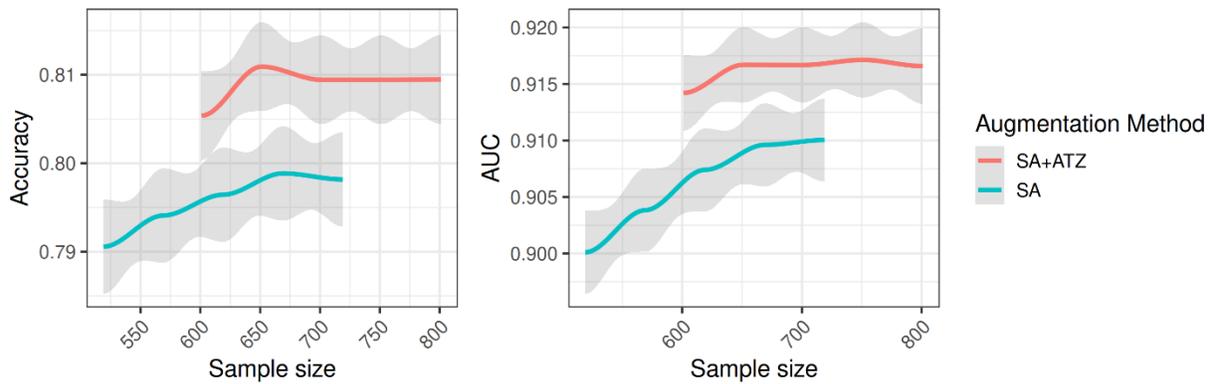

*Figure 5 Accuracy and AUC of the logistic regression fitted on the SURGE-Ahead data (SA) and SURGE-Ahead combined with the ATZ register (SA+ATZ) with additional synthetic data points generated with ARF ranging from 50 to 500. LOESS-smoothed curves (span = 0.3) are fitted to values aggregated over cross-validation replicates and shown across the different sample sizes.*

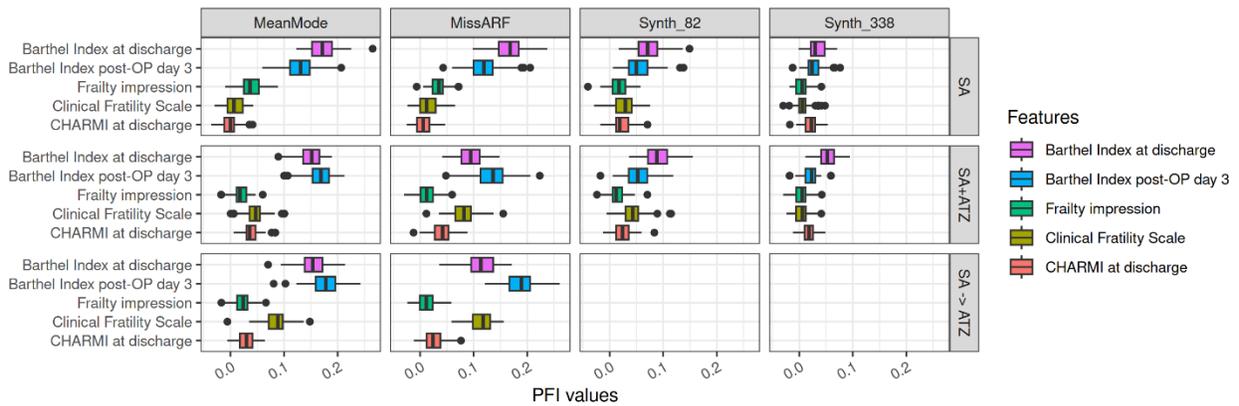

*Figure 6 PFI on test dataset for the logistic regression fitted on SURGE-Ahead data (SA) and its augmented versions of SA. Augmentations include combining SA and the ATZ register with imputation (SA+ATZ, SA → ATZ using MeanMode or MissARF) and synthetic data augmentation with 82 (Synth_82) or 338 (Synth_338) additional samples generated with ARF from SA or SA+ATZ. Shown are the union of the top three highest-ranked features from each setting. The boxplots are plotted over the 100 replicates (each averaged over 5 folds).*

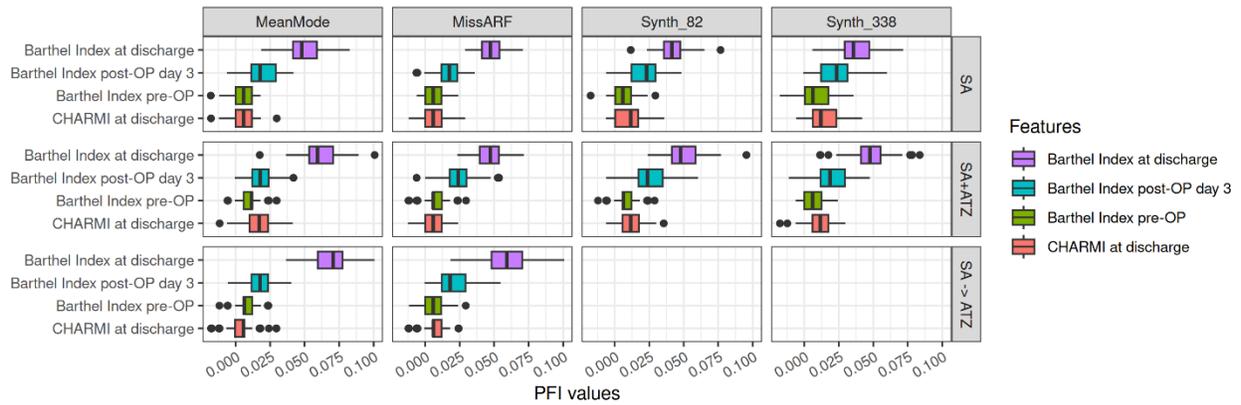

*Figure 7 PFI on test dataset for the random forest fitted on SURGE-Ahead data (SA) and its augmented versions of SA. Augmentations include combining SA and the ATZ register with imputation (SA+ATZ, SA → ATZ using MeanMode or MissARF) and synthetic data augmentation with 82 (Synth_82) or 338 (Synth_338) additional samples generated with ARF from SA or SA+ATZ. Shown are the union of the top three highest-ranked features from each setting. The boxplots are plotted over the 100 replicates (each averaged over 5 folds).*

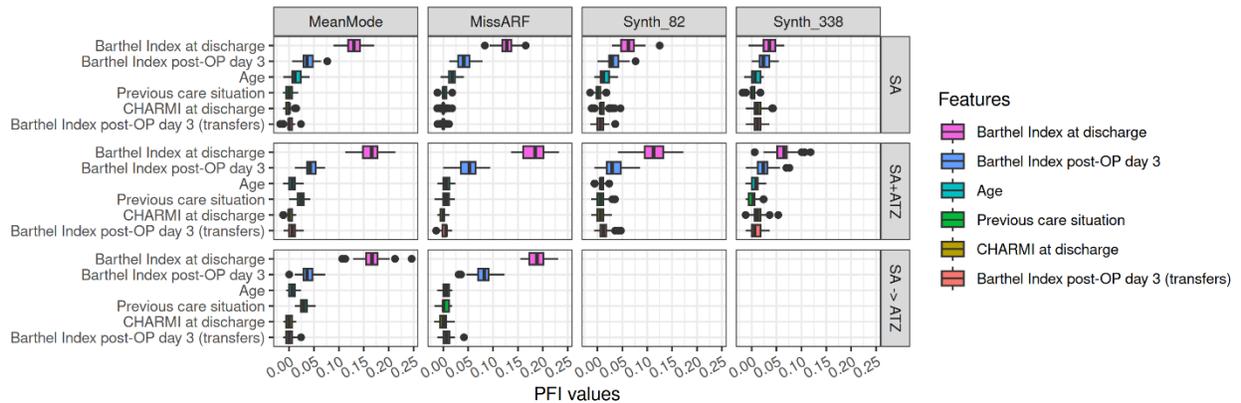

*Figure 8 PFI on test dataset for TabPFN fitted on SURGE-Ahead data (SA) and its augmented versions of SA. Augmentations include combining SA and the ATZ register with imputation (SA+ATZ, SA → ATZ using MeanMode or MissARF) and synthetic data augmentation with 82 (Synth_82) or 338 (Synth_338) additional samples generated with ARF from SA or SA+ATZ. Shown are the union of the top three highest-ranked features from each setting. The boxplots are plotted over the 72 replicates (each averaged over 5 folds); the number of replicates was limited due to higher computational cost.*